\documentclass[prl,aps,twocolumn,showpacs]{revtex4}
\begin{document}
\title{Locality and topology in the molecular Aharonov-Bohm effect}
\author{Erik Sj\"{o}qvist\footnote{Electronic address:
erik.sjoqvist@kvac.uu.se}} 
\affiliation{Department of Quantum Chemistry, Uppsala University, 
Box 518, Se-751 20 Uppsala, Sweden} 
\begin{abstract}
It is shown that the molecular Aharonov-Bohm effect is neither 
nonlocal nor topological in the sense of the standard magnetic 
Aharonov-Bohm effect. It is further argued that there is a 
close relationship between the molecular Aharonov-Bohm effect 
and the Aharonov-Casher effect for an electrically neutral 
spin$-\frac{1}{2}$ particle encircling a line of charge. 
\end{abstract} 
\maketitle
In the standard magnetic Aharonov-Bohm (AB) effect 
\cite{aharonov59}, a charged particle exhibits a 
testable phase shift when encircling a line of magnetic 
flux. The AB effect is nonlocal as it may happen although 
the particle experiences no physical field and no exchange 
of physical quantity takes place along the particle's path 
\cite{peshkin95}. It is topological as it requires the 
particle to be confined to a multiply connected region 
and as any assignment of phase shift along the particle's 
path is necessarily gauge dependent and thus neither 
objective nor experimentally testable \cite{peshkin95}. 

In the molecular Aharonov-Bohm (MAB) effect \cite{mead79}, 
the nuclear motion exhibits a measurable effect under adiabatic 
transport around a conical intersection. A condition for this 
effect to occur is the accumulation along the nuclear path of 
a nontrivial Berry phase \cite{berry84} acquired by the 
corresponding electronic motion. This additional requirement 
makes MAB special and its physical nature potentially different 
from that of the standard AB effect.  

To delineate this difference is the major aim of this Letter. 
We demonstrate that although MAB requires the nuclear motion 
to be confined to a multiply connected region, it fails to obey 
the remaining criteria for a nonlocal and topological effect. 
Thus, it follows from our analysis that MAB should neither 
be regarded as topological nor as nonlocal in the sense of 
the standard AB effect. Instead, MAB displays an adiabatically 
averaged autocorrelation among the electronic variables, which 
in the two-level case resembles that of the local torque on a 
spin$-\frac{1}{2}$ moving in a locally gauge invariant electric 
field. Furthermore, it is possible to relate the MAB effect to 
the noncyclic Berry phase \cite{garcia98} acquired by the 
electronic variables, which is a locally gauge invariant 
quantity that could be tested in polarimetry \cite{garcia98}
or in interferometry \cite{wagh98}. It should be noted that 
an argumentation similar to that of this Letter has been put 
forward in the context of force free electromagnetic effects 
in neutron interferometry \cite{peshkin95}. 

We consider the well studied $E\otimes \epsilon$ Jahn-Teller 
model, in which the symmetry induced degeneracy of two 
electronic states is lifted by their interaction  with a 
doubly degenerate vibrational mode. In the diabatic 
approximation \cite{mead82}, this is described by the 
vibronic Hamiltonian \cite{zwanziger87}
\begin{eqnarray}
H & = & \frac{1}{2} p_{r}^{2} + \frac{1}{2r^{2}} 
p_{\theta}^{2} + \frac{1}{2} r^{2} 
\nonumber \\ 
 & & + k r^{2|\xi|} \Big[ 
\cos (2\xi \theta) \sigma_{x} + \sin (2\xi \theta) \sigma_{y} 
\Big] ,  
\label{eq:vibronicham}
\end{eqnarray}
where $(r,\theta)$ are polar coordinates of the vibrational 
mode, $(p_{r},p_{\theta})$ the corresponding canonical 
momenta, $k$ is the vibronic coupling strength, $\xi = 
\frac{1}{2},-1,...$ describes the order of the effect 
\cite{mixing}, and we have neglected spin-orbit coupling. 
The linear $E\otimes \epsilon$ model is characterized by 
$\xi = \frac{1}{2}$, while in the quadratic model we have 
$\xi = -1$. The electronic degrees of freedom are 
described by Pauli operators defined in terms of the 
diabatic electronic states $|0\rangle$ and $|1\rangle$ as 
$\sigma_{x} = |0 \rangle \langle 1| + |1 \rangle \langle 0|$, 
$\sigma_{y} = -i|0 \rangle \langle 1| + i|1 \rangle \langle 0|$, 
and $\sigma_{z} = |0 \rangle \langle 0| - |1 \rangle \langle 1|$.
The electronic potential energies associated with $H$ are 
$E_{\pm} (r) = \frac{1}{2} r^{2} \pm kr + 1/(8r^{2})$, which, 
by omitting the divergent Born-Huang term $1/(8r^{2})$, conically 
intersect at the origin $r=0$. 

The Born-Oppenheimer regime is attained when the electronic 
potential energies $E_{\pm}$ are well separated. Explicitly 
one may take this to mean that $E_{+}-E_{-}$ is large at the 
minimum of $E_{-}$. For large separation the Born-Huang term 
$1/(8r^{2})$ is negligible so that $E_{-}$ has its minimum 
approximately at $r=k$ yielding the Born-Oppenheimer condition 
$2k^{2} \gg 1$. In this regime the nuclear motion may take 
place on one of the electronic potential energy surfaces, 
accompanied by an additional effective vector potential. 

Similarly, adiabatic motion of the electronic variables 
occurs when \cite{schiff55} $|\langle - |\dot{H}_{e}| + 
\rangle | \ll (E_{+}-E_{-})^{2}$, when treating the 
nuclear degrees of freedom as time-dependent variables. 
Here, the operator $H_{e}= kr^{2|\xi|} [\cos (2\xi \theta) 
\sigma_{x} + \sin (2\xi \theta) \sigma_{y}]$ is the 
electronic Hamiltonian with instantaneous eigenstates 
$|\pm \rangle$ corresponding to the electronic potential 
energies $E_{\pm}$, and we have put $\hbar = 1$. At the 
minimum of $E_{-}$ the adiabaticity condition 
becomes $\xi |\dot{\theta}| \ll 2k^{2}$, where we again 
have neglected the Born-Huang term $1/(8r^{2})$. 

In the Born-Oppenheimer regime consider the nuclear motion on 
the lowest electronic potential energy surface, as described 
by the effective Hamiltonian 
\begin{eqnarray} 
H_{-} = \langle - |H| - \rangle = \frac{1}{2} p_{r}^{2} + 
\frac{1}{2r^{2}} \Big[ p_{\theta} + \xi \Big]^{2} + E_{-} (r)  
\end{eqnarray}
with $|-\rangle$ single-valued around the conical intersection 
at the origin. Here, $\xi {\bf e}_{\theta} /r$ is the MAB vector 
potential that could be absorbed into the phase factor 
\begin{equation} 
\exp \Big[ i\int_{{\bf r}_{0}}^{{\bf r}} \frac{\xi}{r'} 
{\bf e}_{\theta}' \cdot d{\bf r}' \Big] = 
\exp \Big[ i\xi (\theta - \theta_{0}) \Big] , 
\label{eq:mabshift}
\end{equation}
which for the linear case $\xi = \frac{1}{2}$ corresponds to 
a nontrivial sign change for a closed loop only if it encircles 
the conical intersection. This sign change has measurable 
consequences as it restores the original molecular symmetry 
of the vibronic ground state \cite{ham87} and as it shifts 
the spectrum of the quantized nuclear pseudorotation 
\cite{kendrick97}. On the other hand, in the quadratic case 
$\xi = -1$, the phase factor in Eq. (\ref{eq:mabshift}) for 
a closed loop around the conical intersection is $+1$ and 
the MAB vector potential does not have any observable 
consequences on the nuclear motion. 

The electronic Born-Oppenheimer states are eigenstates of 
$\sigma_{x} (\theta) \equiv \cos (2\xi \theta ) \sigma_{x} + 
\sin (2\xi \theta ) \sigma_{y}$. It is therefore perhaps 
tempting to replace the electronic motion by the appropriate 
eigenvalue of $\sigma_{x} (\theta )$ in the Born-Oppenheimer 
regime so that the electronic variables can be ignored, 
creating an illusion that the nontrivial effect of the 
MAB vector potential on the nuclear motion in the $\xi = 
\frac{1}{2}$ case is nonlocal and topological in the sense 
of the standard AB effect. However, the electronic variables 
are dynamical and do not commute among themselves. In particular, 
although the expectation values of the remaining mutually 
complementary electronic observables $\sigma_{y} (\theta ) \equiv - 
\sin (2\xi \theta ) \sigma_{x} + \cos (2\xi \theta ) \sigma_{y}$ 
and $\sigma_{z} (\theta) \equiv \sigma_{z}$ vanish in the 
Born-Oppenheimer limit, their fluctuations do not. As the 
molecule is ideally a closed physical system, its total 
energy is conserved so that equal and opposite fluctuations 
must be exchanged locally with the internal electromagnetic 
field of the molecule during the nuclear pseudorotation. 

To take the argumentation against the nonlocal and topological 
nature of MAB a step further, let us consider the vibronic 
motion in the adiabatic picture. First, we note that the 
motion of $\theta (t)$ depends among other things on the 
motion of the electronic variables. Thus, to distinguish 
the vibronic coupling effect from that associated with the 
dynamics of the nuclei, it turns out to be useful to 
transform the electronic variables to an internal molecular 
frame that co-moves with the pseudorotation. In this frame 
the vibronic Hamiltonian reads 
\begin{equation} 
H' = U^{\dagger}H U = 
\frac{1}{2k^{2}} \big[ p_{\theta} - \xi \sigma_{z} \big]^{2} + 
\frac{k^{2}}{2} + k^{2} \sigma_{x} , 
\end{equation} 
where $U = \exp [ -i \xi \theta \sigma_{z} ]$ is the 
unitary spin rotation operator and we have put $r=k$. 
Using $H'$ and the Heisenberg picture we obtain the 
equations of motion  
\begin{eqnarray}
k^{2} \dot{\theta} & = & p_{\theta} - \xi \sigma_{z} , 
\nonumber \\ 
\dot{p}_{\theta} & = & 0 , 
\nonumber \\ 
\dot{\sigma}_{x} & = & 2\xi \dot{\theta} \sigma_{y}  , 
\nonumber \\  
\dot{\sigma}_{y} & = & - 2\xi \dot{\theta} \sigma_{x} - 
2k^{2} \sigma_{z} , 
\nonumber \\  
\dot{\sigma}_{z} & = & 2k^{2} \sigma_{y} ,  
\label{eq:eqm}
\end{eqnarray} 
where $\xi |\dot{\theta}| \ll 2k^{2}$ in the adiabatic 
limit. It follows that the electronic part describes the 
local torque due to an effective magnetic field ${\bf B} = 
(2k^{2},0,-2\xi \dot{\theta})$ seen by the electronic 
variables in the rotating frame. The large static $x$ 
component of ${\bf B}$ depends only on the vibronic 
coupling parameter $k$ and is irrelevant to MAB. On 
the other hand, the small $z$ component corresponds 
exactly to the MAB effect and gives rise to a 
${\mbox{\boldmath $\sigma$}} \cdot ({\bf v} \times {\bf E})$ 
interaction effect on the spin in its rest frame when it moves 
in the $r-\theta$ plane exposed to the effective electric 
field ${\bf E} = (\xi /r) {\bf e}_{r}$. This effective 
${\bf E}$ field coincides with that of a charged line 
in the $z$ direction sitting at the conical intersection 
and with $\xi$ being proportional to the charge per unit 
length. Thus, the term responsible for MAB resembles exactly 
that of the Aharonov-Casher (AC) effect \cite{aharonov84} 
for an electrically neutral spin$-\frac{1}{2}$ particle 
encircling a line of charge. 

Interpreting the phase shift in Eq. (\ref{eq:mabshift}) 
as an AC effect suggests that MAB is essentially neither 
nonlocal nor topological as it depends on an integral 
whose integrand is proportional to the locally gauge 
invariant effective ${\bf E}$ field and does not depend 
on any physical quantity outside the nuclear path. 
As has been demonstrated in Ref. \cite{peshkin95} in 
the context of force free electromagnetic effects for 
neutrons, the local nature of MAB may be further 
elucidated by considering the relative change of the 
initial and instantaneous electronic variables being 
represented by the vector operators 
${\mbox{\boldmath $\sigma$}}(0)$ and 
${\mbox{\boldmath $\sigma$}} (t)$, respectively. The 
starting point for such a semiclassical analysis is 
to consider in the rotating frame the electronic 
autocorrelation operators 
\begin{eqnarray}
C(t) & = & \frac{1}{4} \Big[ \sigma_{x} (0) \sigma_{x} (t) + 
\sigma_{y} (0) \sigma_{y} (t) + {\text{h.c.}} \Big] ,
\nonumber \\ 
S(t) & = & \frac{1}{4} \Big[ \sigma_{x} (0) \sigma_{y} (t) - 
\sigma_{y} (0) \sigma_{x} (t) + {\text{h.c.}} \Big]  
\end{eqnarray} 
that measure the correlation between the $x-y$ projections 
of ${\mbox{\boldmath $\sigma$}} (0)$ and 
${\mbox{\boldmath $\sigma$}} (t)$. These operators are 
Hermitian and thus measurable in principle. Their equations 
of motion read 
\begin{eqnarray}
\dot{C} & = & 2\xi \dot{\theta} \ S - \xi \dot{\theta} 
\sin (2k^{2}t) , 
\nonumber \\ 
\dot{S} & = & -2\xi \dot{\theta} \ C + \xi \dot{\theta} 
\Big[ 1- \cos (2k^{2}t) \Big]  , 
\label{eq:autoeqm}
\end{eqnarray}
where we have used Eq. (\ref{eq:eqm}). These equations 
are characterized by two time scales: the fast electronic 
oscillations and the slow nuclear pseudorotation, with 
frequencies $2k^{2}$ and $2\xi \dot{\theta}$, respectively. 
Thus, we may simplify Eq. (\ref{eq:autoeqm}) by adiabatic 
averaging \cite{arnold89} over one period of the fast motion 
yielding  
\begin{eqnarray}
\dot{\overline{C}} & = & 2\xi \dot{\theta} \ \overline{S} , 
\nonumber \\ 
\dot{\overline{S}} & = & -2\xi \dot{\theta} \ \overline{C} + 
\xi \dot{\theta} ,  
\end{eqnarray}
which have the solutions 
\begin{eqnarray}
\overline{C} & = & 
\frac{1}{2} \Big( 1 + \cos [2\xi (\theta - \theta_{0})] \Big) , 
\nonumber \\ 
\overline{S} & = & 
- \frac{1}{2} \sin [2\xi (\theta - \theta_{0})] .   
\label{eq:autosolutions}
\end{eqnarray}
This shows that the relative angle $\varphi = 2\xi 
(\theta - \theta_{0})$ between the two $x-y$ projections 
of ${\mbox{\boldmath $\sigma$}} (0)$ and 
${\mbox{\boldmath $\sigma$}} (t)$ is changed by the action 
of the local torque. $\varphi$ is precisely twice the 
MAB phase in Eq. (\ref{eq:mabshift}), where the factor $2$ 
is the usual rotation factor for spin$-\frac{1}{2}$. 
Thus, MAB may be described in terms of the relative angle 
between the $x-y$ projections of ${\mbox{\boldmath $\sigma$}} 
(0)$ and ${\mbox{\boldmath $\sigma$}} (t)$. The change of  
this angle is due to the torque on the electronic variables 
and shows that MAB is essentially a local effect.

There is an objective way to relate the phase shift in Eq. 
(\ref{eq:mabshift}) locally along the nuclear path using 
the noncyclic Berry phase $\gamma_{g}$ of the (lowest) 
electronic Born-Oppenheimer state vector 
\begin{equation} 
|- (\theta) \rangle = \frac{e^{i\alpha (\theta)}}{\sqrt{2}} 
\Big[ e^{-i\xi \theta} |0\rangle + e^{i\xi \theta} |1\rangle 
\Big] . 
\end{equation} 
Here, we assume $\alpha$ to be differentiable along the path 
but otherwise arbitrary. The noncyclic Berry phase is defined 
by removing the accumulation of local phase changes from the 
total phase and is testable in polarimetry \cite{garcia98} or 
in interferometry \cite{wagh98}. We obtain for 
$|- (\theta) \rangle$ \cite{garcia98} 
\begin{eqnarray} 
\gamma_{g} & = & \arg \langle - (\theta_{0}) 
|- (\theta) \rangle + i \int_{\theta_{0}}^{\theta} 
\langle - (\theta') | \frac{\partial}{\partial \theta'} 
|- (\theta') \rangle d\theta' 
\nonumber \\ 
 & = & \arg \cos [\xi (\theta -\theta_{0})] . 
\end{eqnarray}
Clearly, $\gamma_{g}$ is locally gauge invariant as it is 
independent of $\alpha$. It corresponds to a phase jump of 
$\pi$ at $\theta -\theta_{0} = \pi /(2\xi)$, where the overlap 
$\langle - (\theta_{0}) |- (\theta) \rangle$ vanishes. In 
the quadratic $E\times \epsilon$ Jahn-Teller case where 
$\xi = -1$, a closed loop around the conical intersection 
contains two $\pi$ phase jumps, thus explaining the $+1$ 
MAB phase factor. On the other hand, in the linear case 
where $\xi = \frac{1}{2}$, there is only a single $\pi$ 
jump creating a physically nontrivial sign change for such 
a loop. Thus, both the absence in the quadratic case and 
the presence in the linear case of a nontrivial MAB effect 
could be explained locally as they both require the existence 
of points along the nuclear path where the electronic states 
at $\theta_{0}$ and $\theta$ become orthogonal. This local 
assignment of electronic phase shift is gauge invariant at 
each point along the nuclear path and thus experimentally 
testable in principle. It shows that MAB is essentially not 
a topological effect. 

In conclusion, we have shown in the case of the 
$E \otimes \epsilon$ Jahn-Teller model that the molecular 
Aharonov-Bohm (MAB) effect is neither nonlocal nor topological 
in the sense of the standard magnetic Aharonov-Bohm effect.
Locality is preserved as MAB can be explained as a local 
torque on the electronic variables that accumulates along 
the path of the nuclei around the conical intersection. It 
is not topological as it may be described in terms of a 
gauge invariant effective electric field and as there 
is an objective way to relate the phase shift locally along 
the nuclear path via the noncyclic Berry phase. We remark that 
the present analysis also applies to other molecular systems 
that exhibit conical intersections, as well as to the 
microwave resonator experiments recently discussed in 
the literature \cite{lauber94}. 
\vskip 0.5 cm 
This work was supported by the Swedish Research Council. 


\begin{thebibliography}{99}  
\bibitem{aharonov59} Y. Aharonov and D. Bohm, 
Phys. Rev. {\bf 115}, 485 (1959). 
\bibitem{peshkin95} M. Peshkin and H.J. Lipkin, 
Phys. Rev. Lett. {\bf 74}, 2847 (1995). 
\bibitem{mead79} C.A. Mead and D.G. Truhlar, 
J. Chem. Phys. {\bf 70}, 2284 (1979); 
C.A. Mead, 
Chem. Phys. {\bf 49}, 23 (1980); {\it Ibid.} {\bf 49}, 33 (1980).  
\bibitem{berry84} M.V. Berry, 
Proc. Roy. Soc. London Ser. A {\bf 392}, 45 (1984). 
\bibitem{garcia98} G. Garc\'{\i}a de Polavieja and E. Sj\"{o}qvist, 
Am. J. Phys. {\bf 66}, 431 (1998). 
\bibitem{wagh98} A.G. Wagh, V.C. Rakhecha, P. Fischer, and A. Ioffe, 
Phys. Rev. Lett. {\bf 81}, 1992 (1998); 
E. Sj\"{o}qvist, 
Phys. Lett. A {\bf 286}, 4 (2001). 
\bibitem{mead82} C.A. Mead and D.G. Truhlar, 
J. Chem. Phys. {\bf 77}, 6090 (1982). 
\bibitem{zwanziger87} J.W. Zwanziger and E.R. Grant, 
J. Chem. Phys. {\bf 87}, 2954 (1987). 
\bibitem{mixing} For simplicity, it is assumed in Eq. 
(\ref{eq:vibronicham}) that no mixing of different orders $\xi$ 
occurs. Such a mixing introduces new conical intersections, as 
has been shown in Ref. \cite{zwanziger87}, and also changes the 
electronic potential energy surfaces, but the present argumentation 
against the nonlocal and topological nature of MAB would still be 
valid.   
\bibitem{schiff55} L.I. Schiff, 
Quantum mechanics (McGraw-Hill, New York, 1955), 2nd ed., pp. 213-216. 
\bibitem{ham87} F.S. Ham,
Phys. Rev. Lett. {\bf 58}, 725 (1987). 
\bibitem{kendrick97} B. Kendrick, 
Phys. Rev. Lett. {\bf 79}, 2431 (1997); 
H. von Busch, V. Dev, H.-A. Eckel, S. Kasahara, 
J. Wang, W. Demtr\"{o}der, P. Sebald, and W. Meyer, 
Phys. Rev. Lett. {\bf 81}, 4584 (1998).  
\bibitem{aharonov84} Y. Aharonov and A. Casher, 
Phys. Rev. Lett. {\bf 53}, 319 (1984). 
\bibitem{arnold89} V.I. Arnold, Mathematical Methods of Classical 
Mechanics, 2nd Ed. (Springer-Verlag, New York, 1989), ch. 10.
\bibitem{lauber94} H.-M. Lauber, et al. 
Phys. Rev. Lett. {\bf 72}, 1004 (1994); 
D.E. Manolopoulos and M.S. Child, 
Phys. Rev. Lett. {\bf 82}, 2223 (1999); 
F. Pistolesi and N. Manini, 
Phys. Rev. Lett. {\bf 85}, 1585 (2000); 
J. Samuel and A. Dhar, 
Phys. Rev. Lett. {\bf 87}, 260401 (2001). 
\end{thebibliography}
\end{document}